\documentclass[11pt]{cernrep}
\usepackage{graphicx}
\usepackage{cite,./mcite}
\begin{document}


\newcommand{\invfb}{\ensuremath{\mathrm{fb^{-1}}}}
\newcommand{\invpb}{\ensuremath{\mathrm{pb^{-1}}}}
\newcommand{\invnb}{\ensuremath{\mathrm{nb^{-1}}}}
\newcommand{\rhoprim}{\mbox{$\rho^\prime$}}
\newcommand{\tprim}{\mbox{$t^\prime$}}
\newcommand{\rhop}{\mbox{$\rho^\prime$}}
\newcommand{\zet}{\mbox{$\zeta$}}
\newcommand{\rfivecomb}{\mbox{$r^5_{00} + 2 r^5_{11}$}}
\newcommand{\ronecomb}{\mbox{$r^1_{00} + 2 r^1_{11}$}}
\newcommand{\gstarVM} {\mbox{$\gamma^* \ \!p \rightarrow V\ \!Y$}}
\newcommand{\gsrel} {\mbox{$\gamma^* \ \!p \rightarrow \rho \ \!p$}}
\newcommand{\gsrpd} {\mbox{$\gamma^* \ \!p \rightarrow \rho \ \!Y$}}
\newcommand{\gspel} {\mbox{$\gamma^* \ \!p \rightarrow \phi \ \!p$}}
\newcommand{\gsppd} {\mbox{$\gamma^* \ \!p \rightarrow \phi \ \!Y$}}
\newcommand{\gstarp} {\mbox{$\gamma^*\ \!p$}}
\newcommand{\mv} {\mbox{$M_V$}}
\newcommand{\mvsq} {\mbox{$M_V^2$}}
\newcommand{\msq} {\mbox{$M_V^2$}}
\newcommand{\qsqplmsq} {\mbox{($Q^2 \!+ \!M_V^2$})}
\newcommand{\qqsqplmsq} {\mbox{$Q^2 \!+ \!M_V^2}$}
\newcommand{\alprim}{\mbox{$\alpha^\prime$}}
\newcommand{\alphaz}{\mbox{$\alpha(0)$}}
\newcommand{\alpomz}{\mbox{$\alpha_{\PO}(0)$}}
\newcommand{\hence}{\mbox{$=>$}}
\newcommand{\vm}{\mbox{$V\!M$}}
\newcommand{\sur}{\mbox{\ \! / \ \!}}
\newcommand{\tzz} {\mbox{$T_{00}$}}
\newcommand{\tuu} {\mbox{$T_{11}$}}
\newcommand{\tzu} {\mbox{$T_{01}$}}
\newcommand{\tuz} {\mbox{$T_{10}$}}
\newcommand{\tmuu} {\mbox{$T_{-11}$}}
\newcommand{\tumu} {\mbox{$T_{1-1}$}}
\newcommand{\abstzz} {\mbox{$|T_{00}|$}}
\newcommand{\abstuu} {\mbox{$|T_{11}|$}}
\newcommand{\abstzu} {\mbox{$|T_{01}|$}}
\newcommand{\abstuz} {\mbox{$|T_{10}|$}}
\newcommand{\abstmuu} {\mbox{$|T_{-11}|$}}
\newcommand{\ralpha} {\mbox{$\tuu \sur \tzz$}}
\newcommand{\rbeta} {\mbox{$\tzu \sur \tzz$}}
\newcommand{\rdelta} {\mbox{$\tuz \sur \tzz$}}
\newcommand{\reta} {\mbox{$\tmuu \sur \tzz$}}
\newcommand{\averm} {\mbox{$\av {M}$}}
\newcommand{\rapproch} {\mbox{$R_{SCHC+T_{01}}$}}
\newcommand{\chisq} {\mbox{$\chi^2 / {\rm d.o.f.}$}}


%
%
\newcommand{\s}{\mbox{$s$}}
\newcommand{\ttra}{\mbox{$t$}}
\newcommand{\modt}{\mbox{$|t|$}}
\newcommand{\eminpz}{\mbox{$E-p_z$}}
\newcommand{\eminpzs}{\mbox{$\Sigma(E-p_z)$}}
\newcommand{\rap}{\ensuremath{\eta^*} }
\newcommand{\W}{\mbox{$W$}}
\newcommand{\w}{\mbox{$W$}}
\newcommand{\Q}{\mbox{$Q$}}
\newcommand{\q}{\mbox{$Q$}}
\newcommand{\xB}{\mbox{$x$}}  
\newcommand{\xF}{\mbox{$x_F$}}  
\newcommand{\xg}{\mbox{$x_g$}}  
\newcommand{\xbj}{x}
\newcommand{\xpom}{x_{\PO}}
\newcommand{\y}{\mbox{$y~$}}
\newcommand{\Qsq}{\mbox{$Q^2$}}
\newcommand{\qsq}{\mbox{$Q^2$}}
\newcommand{\kjet}{\mbox{$k_{T\rm{jet}}$}}
\newcommand{\xjet}{\mbox{$x_{\rm{jet}}$}}
\newcommand{\Ejet}{\mbox{$E_{\rm{jet}}$}}
\newcommand{\thjet}{\mbox{$\theta_{\rm{jet}}$}}
\newcommand{\pjet}{\mbox{$p_{T\rm{jet}}$}}
\newcommand{\et}{\mbox{$E_T~$}}
\newcommand{\kt}{\mbox{$k_T~$}}
\newcommand{\ptrans}{\mbox{$p_T~$}}
\newcommand{\pth}{\mbox{$p_T^h~$}}
\newcommand{\pte}{\mbox{$p_T^e~$}}
\newcommand{\ptsq}{\mbox{$p_T^{\star 2}~$}}
\newcommand{\as}{\mbox{$\alpha_s~$}}
\newcommand{\ycut}{\mbox{$y_{\rm cut}~$}}
\newcommand{\gx}{\mbox{$g(x_g,Q^2)$~}}
\newcommand{\xpart}{\mbox{$x_{\rm part~}$}}
\newcommand{\mrsdm}{\mbox{${\rm MRSD}^-~$}}
\newcommand{\mrsdmp}{\mbox{${\rm MRSD}^{-'}~$}}
\newcommand{\mrsdn}{\mbox{${\rm MRSD}^0~$}}
\newcommand{\lambdams}{\mbox{$\Lambda_{\rm \bar{MS}}~$}}
%
%
\newcommand{\gp}{\ensuremath{\gamma}p }
\newcommand{\gammasp}{\ensuremath{\gamma}*p }
\newcommand{\gammap}{\ensuremath{\gamma}p }
\newcommand{\gsp}{\ensuremath{\gamma^*}p }
\newcommand{\dsiget}{\ensuremath{{\rm d}\sigma_{ep}/{\rm d}E_t^*} }
\newcommand{\dsigrap}{\ensuremath{{\rm d}\sigma_{ep}/{\rm d}\eta^*} }
\newcommand{\epem}{\mbox{$e^+e^-$}}
\newcommand{\ep}{\mbox{$ep~$}}
\newcommand{\epl}{\mbox{$e^{+}$}}
\newcommand{\emi}{\mbox{$e^{-}$}}
\newcommand{\epm}{\mbox{$e^{\pm}$}}
\newcommand{\se}{section efficace}
\newcommand{\ses}{sections efficaces}
%
%
\newcommand{\phib}{\mbox{$\varphi$}}
\newcommand{\rh}{\mbox{$\rho$}}
\newcommand{\rhz}{\mbox{$\rh^0$}}
\newcommand{\ph}{\mbox{$\phi$}}
\newcommand{\om}{\mbox{$\omega$}}
\newcommand{\jpsi}{\mbox{$J/\psi$}}
\newcommand{\pipi}{\mbox{$\pi^+\pi^-$}}
\newcommand{\pip}{\mbox{$\pi^+$}}
\newcommand{\pim}{\mbox{$\pi^-$}}
\newcommand{\kk}{\mbox{K^+K^-$}}
\newcommand{\bsl}{\mbox{$b$}}
\newcommand{\alp}{\mbox{$\alpha^\prime$}}
\newcommand{\alpom}{\mbox{$\alpha_{\PO}$}}
\newcommand{\alpomp}{\mbox{$\alpha_{\PO}^\prime$}}
\newcommand{\rzzzz}{\mbox{$r_{00}^{04}$}}
\newcommand{\rzqzz}{\mbox{$r_{00}^{04}$}}
\newcommand{\rzquz}{\mbox{$r_{10}^{04}$}}
\newcommand{\rzqumu}{\mbox{$r_{1-1}^{04}$}}
\newcommand{\ruuu}{\mbox{$r_{11}^{1}$}}
\newcommand{\ruzz}{\mbox{$r_{00}^{1}$}}
\newcommand{\ruuz}{\mbox{$r_{10}^{1}$}}
\newcommand{\ruumu}{\mbox{$r_{1-1}^{1}$}}
\newcommand{\rduz}{\mbox{$r_{10}^{2}$}}
\newcommand{\rdumu}{\mbox{$r_{1-1}^{2}$}}
\newcommand{\rcuu}{\mbox{$r_{11}^{5}$}}
\newcommand{\rczz}{\mbox{$r_{00}^{5}$}}
\newcommand{\rcuz}{\mbox{$r_{10}^{5}$}}
\newcommand{\rcumu}{\mbox{$r_{1-1}^{5}$}}
\newcommand{\rsuz}{\mbox{$r_{10}^{6}$}}
\newcommand{\rsumu}{\mbox{$r_{1-1}^{6}$}}
\newcommand{\rzqik}{\mbox{$r_{ik}^{04}$}}
\newcommand{\rhzik}{\mbox{$\rh_{ik}^{0}$}}
\newcommand{\rhqik}{\mbox{$\rh_{ik}^{4}$}}
\newcommand{\rhaik}{\mbox{$\rh_{ik}^{\alpha}$}}
\newcommand{\rhzzz}{\mbox{$\rh_{00}^{0}$}}
\newcommand{\rhqzz}{\mbox{$\rh_{00}^{4}$}}
\newcommand{\raik}{\mbox{$r_{ik}^{\alpha}$}}
\newcommand{\razz}{\mbox{$r_{00}^{\alpha}$}}
\newcommand{\rauz}{\mbox{$r_{10}^{\alpha}$}}
\newcommand{\raumu}{\mbox{$r_{1-1}^{\alpha}$}}

\newcommand{\R}{\mbox{$R$}}
\newcommand{\rzero}{\mbox{$r_{00}^{04}$}}
\newcommand{\rone}{\mbox{$r_{1-1}^{1}$}}
\newcommand{\costh}{\mbox{$\cos\theta$}}
\newcommand{\cosp}{\mbox{$\cos\psi$}}
\newcommand{\costop}{\mbox{$\cos(2\psi)$}}
\newcommand{\cosd}{\mbox{$\cos\delta$}}
\newcommand{\cossqp}{\mbox{$\cos^2\psi$}}
\newcommand{\cossqt}{\mbox{$\cos^2\theta^*$}}
\newcommand{\sint}{\mbox{$\sin\theta^*$}}
\newcommand{\sintot}{\mbox{$\sin(2\theta^*)$}}
\newcommand{\sinsqt}{\mbox{$\sin^2\theta^*$}}
\newcommand{\costhst}{\mbox{$\cos\theta^*$}}
\newcommand{\vep}{\mbox{$V p$}}
\newcommand{\mpipi}{\mbox{$m_{\pi^+\pi^-}$}}
\newcommand{\mkk}{\mbox{$m_{KK}$}}
\newcommand{\mkaka}{\mbox{$m_{K^+K^-}$}}
\newcommand{\mpp}{\mbox{$m_{\pi\pi}$}}       
\newcommand{\mppsq}{\mbox{$m_{\pi\pi}^2$}}   
\newcommand{\mpi}{\mbox{$m_{\pi}$}}          
\newcommand{\mrho}{\mbox{$m_{\rho}$}}        
\newcommand{\mrhosq}{\mbox{$m_{\rho}^2$}}    
\newcommand{\Gmpp}{\mbox{$\Gamma (\mpp)$}}   
\newcommand{\Gmppsq}{\mbox{$\Gamma^2(\mpp)$}}
\newcommand{\Grho}{\mbox{$\Gamma_{\rho}$}}   
\newcommand{\grho}{\mbox{$\Gamma_{\rho}$}}   
\newcommand{\Grhosq}{\mbox{$\Gamma_{\rho}^2$}}   
%
%
\newcommand{\cm}{\mbox{\rm cm}}
\newcommand{\GeV}{\mbox{\rm GeV}}
\newcommand{\gev}{\mbox{\rm GeV}}
\newcommand{\GeVx}{\rm GeV}
\newcommand{\gevx}{\rm GeV}
\newcommand{\GeVc}{\rm GeV/c}
\newcommand{\gevc}{\rm GeV/c}
\newcommand{\MeVc}{\rm MeV/c}
\newcommand{\mevc}{\rm MeV/c}
\newcommand{\MeV}{\mbox{\rm MeV}}
\newcommand{\mev}{\mbox{\rm MeV}}
\newcommand{\MeVx}{\mbox{\rm MeV}}
\newcommand{\mevx}{\mbox{\rm MeV}}
\newcommand{\GeVsq}{\mbox{${\rm GeV}^2$}}
\newcommand{\gevsq}{\mbox{${\rm GeV}^2$}}
\newcommand{\gevsqc}{\mbox{${\rm GeV^2/c^4}$}}
\newcommand{\gevcsq}{\mbox{${\rm GeV/c^2}$}}
\newcommand{\mevcsq}{\mbox{${\rm MeV/c^2}$}}
\newcommand{\GeVsqm}{\mbox{${\rm GeV}^{-2}$}}
\newcommand{\gevsqm}{\mbox{${\rm GeV}^{-2}$}}
\newcommand{\nb}{\mbox{${\rm nb}$}}
\newcommand{\nbinv}{\mbox{${\rm nb^{-1}}$}}
\newcommand{\pbinv}{\mbox{${\rm pb^{-1}}$}}
\newcommand{\mm}{\mbox{$\cdot 10^{-2}$}}
\newcommand{\mmm}{\mbox{$\cdot 10^{-3}$}}
\newcommand{\mmmm}{\mbox{$\cdot 10^{-4}$}}
\newcommand{\degr}{\mbox{$^{\circ}$}}
%
%
\newcommand{\F}{$ F_{2}(x,Q^2)\,$}  
\newcommand{\Fc}{$ F_{2}\,$}    
\newcommand{\XP}{x_{{I\!\!P}/{p}}}       
\newcommand{\TOSS}{x_{{i}/{\PO}}}        
\newcommand{\un}[1]{\mbox{\rm #1}} 
\newcommand{\LO}{Leading Order}
\newcommand{\NLO}{Next to Leading Order}
\newcommand{\ft}{$ F_{2}\,$}
%
%
\newcommand{\mc}{\multicolumn}
\newcommand{\bce}{\begin{center}}
\newcommand{\ece}{\end{center}}
\newcommand{\beq}{\begin{equation}}
\newcommand{\eeq}{\end{equation}}
\newcommand{\bea}{\begin{eqnarray}}
\newcommand{\eea}{\end{eqnarray}}
%
%
\def\lsim{\mathrel{\rlap{\lower4pt\hbox{\hskip1pt$\sim$}}
    \raise1pt\hbox{$<$}}}         
\def\gsim{\mathrel{\rlap{\lower4pt\hbox{\hskip1pt$\sim$}}
    \raise1pt\hbox{$>$}}}         
%
%
\newcommand{\pom}{{I\!\!P}}
\newcommand{\PO}{I\!\!P}
\newcommand{\slowpi}{\pi_{\mathit{slow}}}
\newcommand{\fiidiii}{F_2^{D(3)}}
\newcommand{\fiidiiiarg}{\fiidiii\,(\beta,\,Q^2,\,x)}
\newcommand{\n}{1.19\pm 0.06 (stat.) \pm0.07 (syst.)}
\newcommand{\nz}{1.30\pm 0.08 (stat.)^{+0.08}_{-0.14} (syst.)}
\newcommand{\fiidiiiful}{F_2^{D(4)}\,(\beta,\,Q^2,\,x,\,t)}
\newcommand{\fiipom}{\tilde F_2^D}
\newcommand{\ALPHA}{1.10\pm0.03 (stat.) \pm0.04 (syst.)}
\newcommand{\ALPHAZ}{1.15\pm0.04 (stat.)^{+0.04}_{-0.07} (syst.)}
\newcommand{\fiipomarg}{\fiipom\,(\beta,\,Q^2)}
\newcommand{\pomflux}{f_{\pom / p}}
\newcommand{\nxpom}{1.19\pm 0.06 (stat.) \pm0.07 (syst.)}
\newcommand {\gapprox}
   {\raisebox{-0.7ex}{$\stackrel {\textstyle>}{\sim}$}}
\newcommand {\lapprox}
   {\raisebox{-0.7ex}{$\stackrel {\textstyle<}{\sim}$}}
\newcommand{\pomfluxarg}{f_{\pom / p}\,(x_\pom)}
\newcommand{\dsf}{\mbox{$F_2^{D(3)}$}}
\newcommand{\dsfva}{\mbox{$F_2^{D(3)}(\beta,Q^2,x_{I\!\!P})$}}
\newcommand{\dsfvb}{\mbox{$F_2^{D(3)}(\beta,Q^2,x)$}}
\newcommand{\dsfpom}{$F_2^{I\!\!P}$}
\newcommand{\gap}{\stackrel{>}{\sim}}
\newcommand{\lap}{\stackrel{<}{\sim}}
\newcommand{\fem}{$F_2^{em}$}
\newcommand{\tsnmp}{$\tilde{\sigma}_{NC}(e^{\mp})$}
\newcommand{\tsnm}{$\tilde{\sigma}_{NC}(e^-)$}
\newcommand{\tsnp}{$\tilde{\sigma}_{NC}(e^+)$}
\newcommand{\st}{$\star$}
\newcommand{\sst}{$\star \star$}
\newcommand{\ssst}{$\star \star \star$}
\newcommand{\sssst}{$\star \star \star \star$}
\newcommand{\tw}{\theta_W}
\newcommand{\sw}{\sin{\theta_W}}
\newcommand{\cw}{\cos{\theta_W}}
\newcommand{\sww}{\sin^2{\theta_W}}
\newcommand{\cww}{\cos^2{\theta_W}}
\newcommand{\trm}{m_{\perp}}
\newcommand{\trp}{p_{\perp}}
\newcommand{\trmm}{m_{\perp}^2}
\newcommand{\trpp}{p_{\perp}^2}
\newcommand{\ev}{\'ev\'enement}
\newcommand{\evs}{\'ev\'enements}
\newcommand{\mdv}{mod\`ele \`a dominance m\'esovectorielle}
\newcommand{\mdmv}{mod\`ele \`a dominance m\'esovectorielle}
\newcommand{\mdm}{mod\`ele \`a dominance m\'esovectorielle}
\newcommand{\idiff}{interaction diffractive}
\newcommand{\idiffs}{interactions diffractives}
\newcommand{\pdmv}{production diffractive de m\'esons vecteurs}
\newcommand{\pdmr}{production diffractive de m\'esons \rh}
\newcommand{\pdmp}{production diffractive de m\'esons \ph}
\newcommand{\pdmo}{production diffractive de m\'esons \om}
\newcommand{\pdm}{production diffractive de m\'esons}
\newcommand{\pdiff}{production diffractive}
\newcommand{\diff}{diffractive}
\newcommand{\produ}{production}
\newcommand{\mvs}{m\'esons vecteurs}
\newcommand{\me}{m\'eson}
\newcommand{\mr}{m\'eson \rh}
\newcommand{\mph}{m\'eson \ph}
\newcommand{\mo}{m\'eson \om}
\newcommand{\mrs}{m\'esons \rh}
\newcommand{\mps}{m\'esons \ph}
\newcommand{\mos}{m\'esons \om}
\newcommand{\photo}{photoproduction}
\newcommand{\agq}{\`a grand \qsq}
\newcommand{\agqsq}{\`a grand \qsq}
\newcommand{\apq}{\`a petit \qsq}
\newcommand{\apqsq}{\`a petit \qsq}
\newcommand{\de}{d\'etecteur}
%
%
\newcommand{\sqrts}{$\sqrt{s}$}
\newcommand{\Oa}{$O(\alpha_s)$}
\newcommand{\Oaa}{$O(\alpha_s^2)$}
\newcommand{\PT}{p_{\perp}}
\newcommand{\sh}{\hat{s}}
\newcommand{\uh}{\hat{u}}
\newcommand{\ttbs}{\char'134}
\newcommand{\xpomlo}{3\times10^{-4}}
\newcommand{\xpomup}{0.05}
\newcommand{\llq}{$\alpha_s \ln{(\qsq / \Lambda_{QCD}^2)}$}
\newcommand{\llqx}{$\alpha_s \ln{(\qsq / \Lambda_{QCD}^2)} \ln{(1/x)}$}
\newcommand{\llx}{$\alpha_s \ln{(1/x)}$}
%
%
\newcommand{\Brodsky}{Brodsky {\it et al.}}
\newcommand{\FKS}{Frankfurt, Koepf and Strikman}
\newcommand{\Kop}{Kopeliovich {\it et al.}}
\newcommand{\Ginzburg}{Ginzburg {\it et al.}}
\newcommand{\Ryskin}{\mbox{Ryskin}}
\newcommand{\Kaidalov}{Kaidalov {\it et al.}}
%
%
\def\ar#1#2#3   {{\em Ann. Rev. Nucl. Part. Sci.} {\bf#1} (#2) #3}
\def\epj#1#2#3  {{\em Eur. Phys. J.} {\bf#1} (#2) #3}
\def\err#1#2#3  {{\it Erratum} {\bf#1} (#2) #3}
\def\ib#1#2#3   {{\it ibid.} {\bf#1} (#2) #3}
\def\ijmp#1#2#3 {{\em Int. J. Mod. Phys.} {\bf#1} (#2) #3}
\def\jetp#1#2#3 {{\em JETP Lett.} {\bf#1} (#2) #3}
\def\mpl#1#2#3  {{\em Mod. Phys. Lett.} {\bf#1} (#2) #3}
\def\nim#1#2#3  {{\em Nucl. Instr. Meth.} {\bf#1} (#2) #3}
\def\nc#1#2#3   {{\em Nuovo Cim.} {\bf#1} (#2) #3}
\def\np#1#2#3   {{\em Nucl. Phys.} {\bf#1} (#2) #3}
\def\pl#1#2#3   {{\em Phys. Lett.} {\bf#1} (#2) #3}
\def\prep#1#2#3 {{\em Phys. Rep.} {\bf#1} (#2) #3}
\def\prev#1#2#3 {{\em Phys. Rev.} {\bf#1} (#2) #3}
\def\prl#1#2#3  {{\em Phys. Rev. Lett.} {\bf#1} (#2) #3}
\def\ptp#1#2#3  {{\em Prog. Th. Phys.} {\bf#1} (#2) #3}
\def\rmp#1#2#3  {{\em Rev. Mod. Phys.} {\bf#1} (#2) #3}
\def\rpp#1#2#3  {{\em Rep. Prog. Phys.} {\bf#1} (#2) #3}
\def\sjnp#1#2#3 {{\em Sov. J. Nucl. Phys.} {\bf#1} (#2) #3}
\def\spj#1#2#3  {{\em Sov. Phys. JEPT} {\bf#1} (#2) #3}
\def\zp#1#2#3   {{\em Zeit. Phys.} {\bf#1} (#2) #3}
%
%
\newcommand{\clearemptydoublepage}{\newpage{\pagestyle{empty}\cleardoublepage}}
\newcommand{\scaption}[1]{\caption{\protect{\footnotesize  #1}}}
\newcommand{\proc}[2]{\mbox{$ #1 \rightarrow #2 $}}
\newcommand{\average}[1]{\mbox{$ \langle #1 \rangle $}}
\newcommand{\av}[1]{\mbox{$ \langle #1 \rangle $}}



\title{Exclusive Vector Meson Production and Deeply Virtual Compton Scattering at 
HERA~\footnote{To be published in the Proceedings of the HERA - LHC Workshop.}   }

\author{Alessia Bruni$^1$, Xavier Janssen$^2$, Pierre Marage$^2$}
\institute{
$^1$ Istituto Nazionale di Fisica Nucleare, Via Irnerio 46, I-40126 Bologna,
Italy\\
$^2$ Faculty of Science, Universit\'e Libre de Bruxelles, Bd. du Triomphe, 
B-1040 Brussels, Belgium}
\maketitle

\begin{abstract}
Exclusive vector meson production and deeply virtual Compton scattering are ideally suited reactions 
for studying the structure of the proton and the transition from soft to hard processes.
The main experimental data obtained at HERA are summarised and presented in 
the light of QCD approaches.
\end{abstract}

\section{Introduction}
                                                     \label{sec:intro}

\begin{figure}[h]
\includegraphics[width=0.25\columnwidth]{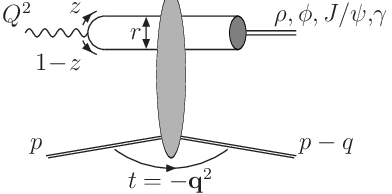}
\includegraphics[width=0.02\columnwidth]{fig/whitebox.eps}
\includegraphics[width=0.21\columnwidth]{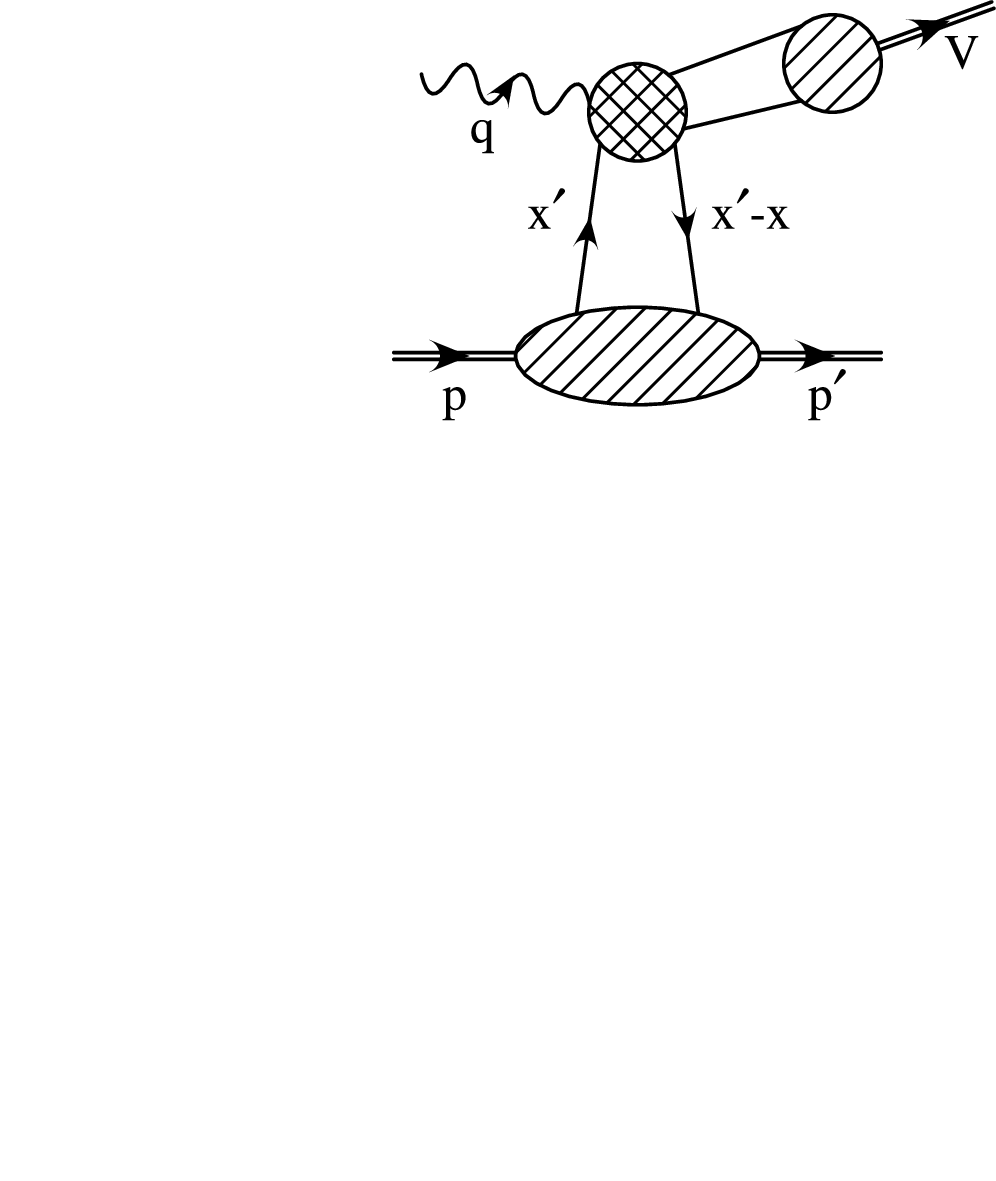}
\includegraphics[width=0.02\columnwidth]{fig/whitebox.eps}
\includegraphics[width=0.45\columnwidth]{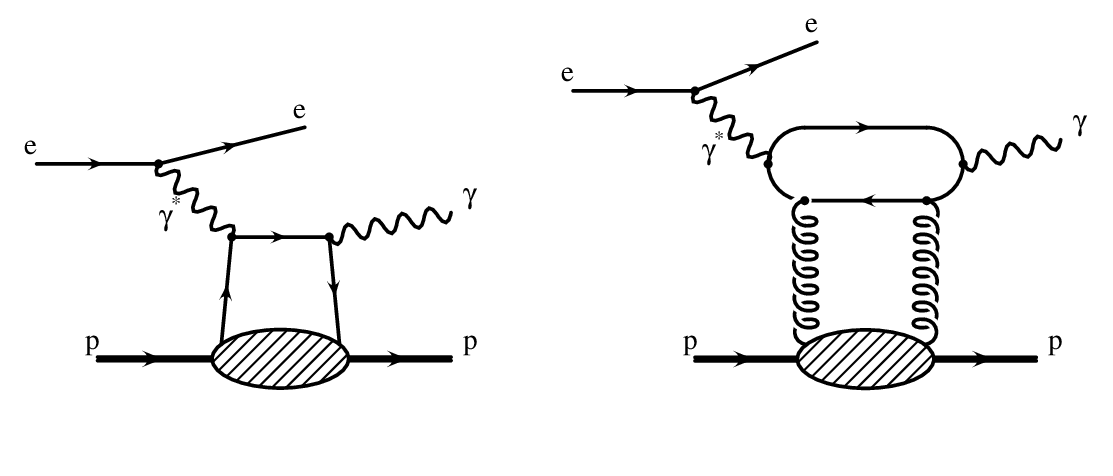} 
\caption{
(from left to right) Representative diagrams of a) low $x$ dipole approach and b) GPD approach,
for VM production; 
c) LO scattering and d) two gluon exchange, for the DVCS process.}
\label{fig:VM}
\end{figure}

The two processes which are the object of the present report,
the exclusive production of a vector meson (VM) of mass $M_V$, $e + p \to e + VM + Y$, and 
deeply virtual Compton scattering (DVCS), $e + p \to e + \gamma + Y$,
where $Y$ is a proton (elastic scattering) or a diffractively excited system 
(proton dissociation), are characterised in Fig.~\ref{fig:VM}.
The kinematical variables are \qsq, the negative square of the photon four-momentum, $W$ the 
photon-proton centre of mass energy ($W^2 \simeq \qsq\ (1 / x - 1)$, $x$ being the Bjorken 
scaling variable) 
and $t$, the square of the four-momentum transfer at the proton vertex.

The H1 and ZEUS collaborations at HERA have studied the elastic and proton dissociative 
production of 
$\rho$~\cite{h1rhophp,*Derrick:1996vw,*Breitweg:1997ed,*H1prelim-06-011,
*Breitweg:1998nh,*Breitweg:1999fm,*h1rho2000,*h1rho2002,Chekanov:2007zr,
janssen2008,*janssen2008-1,hight_zeus,*Aktas:2006qs}, 
$\omega$~\cite{Derrick:1996yt,*Breitweg:2000mu},
$\phi$~\cite{Derrick:1996af,*h1phi2000,*Chekanov:2005cqa,janssen2008}, 
$J/\psi$~\cite{Chekanov:2002xi,*Chekanov:2004mw,*Aktas:2005xu,Aktas:2003zi,*hight_jpsi_zeus}, 
$\psi(2s)$~\cite{Adloff:2002re} 
and $\Upsilon$~\cite{Breitweg:1998ki,*upsilon_h1,upsilon-zeus07}
mesons, and the DVCS process in the elastic 
channel~\cite{Chekanov:2003ya,*Aktas:2005ty,Aaron:2007cz}.
The measurements are performed in the low $x$, large $W$ domain
$10^{-4}\ \lapprox\ x\ \lapprox\ 10^{-2}$, $30 \leq W \leq 300~\gev$.
They cover photoproduction ($\qsq \simeq 0$), with \modt\ values up to 30~\gevsq,
and electroproduction in the deep inelastic (DIS) domain 
($2\ \leq Q^2 \leq 90~\gevsq$) with $\modt\ \lapprox\ 2~\gevsq$.
The cross sections, expressed in terms of $\gamma^*p$ scattering, are measured 
differentially in $Q^2$, $W$ and $t$. 
The measurement of angular distributions gives access to spin density matrix elements 
and polarised amplitudes.

\subsection{Production mechanisms}
                                                      \label{sec:VM}

Within the QCD formalism, two main complementary approaches are used to describe VM 
production and DVCS: dipole factorisation and collinear factorisation. 

\paragraph{Dipole approach of VM production}
At high energy, i.e. small $x$, VM production can be described in the proton rest frame with
three factorising contributions~\cite{Mueller:1989st,*Nikolaev:1990ja} (see Fig.~\ref{fig:VM}a):
the fluctuation of the virtual photon into a $q \bar q$ colour dipole, 
the elastic or proton dissociative dipole--proton scattering, 
and the $q \bar q$ recombination into the final state VM.
The dipole--proton cross section is expected to be flavour independent and governed by the 
transverse size of the dipole.
Light VM photoproduction is dominated by large dipoles, leading to large interaction cross 
sections with the incoming proton, similar to soft hadron--hadron interactions.
In contrast, heavy VM production and large \qsq\ processes are dominated by small dipoles,
with smaller cross sections implied in QCD by colour transparency, the quark and the
antiquark separated by a small distance tending to screen each other's colour. 

The cross section for VM production can be computed at small $x$ and for all \qsq\ values
through models~\cite{Forshaw:2003ki,Kowalski:2006hc,Dosch:2006kz}
using universal dipole--proton cross sections measured in inclusive processes, 
possibly including saturation 
effects~\cite{GolecBiernat:1998js,*GolecBiernat:1999qd,*Munier:2001nr,*Iancu:2003ge}
(see also~\cite{motyka}).
This formalism thus connects the inclusive and diffractive cross sections, also in the
absence of a hard scale. 

In the presence of a hard scale (large quark mass or $Q$), the dipole--proton scattering is 
modelled in perturbative QCD (pQCD) as the exchange of a colour singlet system consisting of
a gluon pair (at lowest order) or a BFKL ladder (at leading logarithm approximation, 
LL $1/x$).
At these approximations, the cross sections are proportional to the square of the gluon density 
$|xG(x)|^2$ in the proton~\cite{Ryskin:1992ui,*Brodsky:1994kf}.
The pQCD
calculations~\cite{Frankfurt:1995jw,Ivanov:2004ax,Martin:1996bp,Ivanov:1998gk}
use $k_t$-unintegrated gluon distributions (see also~\cite{teubner}).
The typical interaction scale is $\mu^2  \simeq z (1-z) (Q^2 + M_V^2)$, where $z$ is 
the fraction of the photon longitudinal momentum carried by the quark.
For heavy VM (in the non-relativistic wave function (WF) approximation) and for light VM
production from longitudinally polarised photons, $z \simeq 1/2$ and the cross sections are 
expected to scale with the variable $\mu^2 = \qsqplmsq / 4$.
In contrast, for light VM production by transversely polarised photons, contributions with 
$z \to 0, 1$ result in the presence of large dipoles and the damping of the scale $\mu$, 
thus introducing  non-perturbative features even for non-small \qsq.

\paragraph{Collinear factorisation and GPD}
In a complementary approach (see Fig.~\ref{fig:VM}b), a collinear factorisation 
theorem~\cite{Collins:1996fb} has been proven in QCD for longitudinal amplitudes in the 
DIS domain, which does not require low $x$ values.
This allows separating contributions from different scales, a large scale at the photon 
vertex, provided by the photon virtuality $Q$ (or the quark mass), and a
small scale for the proton structure.
The latter is described by Generalised Parton Distributions 
(GPD -- see e.g. the reviews~\cite{Diehl:2003ny,*Belitsky:2005qn}), which 
take into account the distribution of transverse momenta of partons with respect to the 
proton direction and longitudinal momentum correlations between partons.
They account for ``off-diagonal" or ``skewing" effects arising from 
the kinematic matching between the initial state (virtual) photon and the final state, VM 
or real photon for DVCS.
GPD calculations have been performed for light VM electroproduction~\cite{kroll}.
NLO corrections to light VM electroproduction and to heavy VM photoproduction have been
computed~\cite{Ivanov:2004zv,*Ivanov:2004vd,*Ivanov:2007je}.


\paragraph{DVCS}
Following collinear factorisation, the DVCS process is described at LO by 
Fig.~\ref{fig:VM}c, where the virtual photon couples directly to a quark in the proton.
QCD calculations at the scale $\mu^2 = \qsq$ involve GPD 
distributions~\cite{Freund:2003qs,Favart:2003cu}.
At higher order, two gluon exchange as in Fig.~\ref{fig:VM}d gives also an important 
contribution at HERA.
Joint fits to DVCS and inclusive structure functions data have been used to extract 
GPD distributions~\cite{Kumericki:2007sa}.

\paragraph{Large \boldmath {\modt} production}
Calculations for VM production at large \modt\ have been performed both in a DGLAP and in
a BFKL approach (see section~\ref{sec:large t}).

\subsection{Measurements at HERA}
                                                      \label{sec:measurements}


Vector mesons are identified by H1 and ZEUS via their decay to two oppositely charged 
particles $\rho \rightarrow \pi^+\pi^-$, $\phi\rightarrow K^+ K^-$, 
$J/\psi \rightarrow e^+e^-, \mu^+\mu^-$ and $\Upsilon \rightarrow \mu^+\mu^-$.
The kinematic 
variables are reconstructed from the scattered electron and 
decay particle measurements.
Forward calorimeters and taggers at small angles are generally used to separate elastic and
proton dissociative events.
The scattered proton is also measured in forward proton spectrometers, with an acceptance
of a few \%,
allowing the selection of a purely elastic sample and the direct measurement of the $t$ variable.

VM production has been investigated  mainly using the HERA I data, collected 
between 1992 and 2000 and  corresponding to an integrated luminosity of 
$\simeq 150$ pb$^{-1}$ for both collaborations.
The integrated luminosity of 500 pb$^{-1}$ collected at HERA II (2003-2007) 
has been analysed so far for DVCS~\cite{Aaron:2007cz} and 
$\Upsilon$~\cite{upsilon-zeus07}.
For HERA II, ZEUS has installed a microvertex detector but has removed the small angle 
detectors: the leading proton spectrometer and the forward and rear calorimeters, 
compromising the precise analysis of diffractive data.
The HERA II analyses of H1 will benefit of the fast track trigger installed in 2002 and, for
general diffraction studies, of the very forward proton spectrometer VFPS 
installed in 2003, which however has very limited acceptance for VM.

\section{From soft to hard diffraction: {\boldmath $t$} dependences and the size of the interaction}
                                                      \label{sec:t}

The $t$ dependences of DVCS and VM production provide information on the size and the 
dynamics of the processes and on the scales relevant for the dominance of perturbative, hard effects.
Whereas total cross sections ($F_2$  measurements) are related, through the optical theorem, 
to the scattering amplitudes in the forward direction, diffractive final states 
provide a unique opportunity to study the region of non-zero momentum transfer $t$.
This gives indirect information on the variable conjugate to $t$, the transverse size of the interaction.

\begin{figure}[htbp]
\begin{center}
\includegraphics[width=0.40\columnwidth]{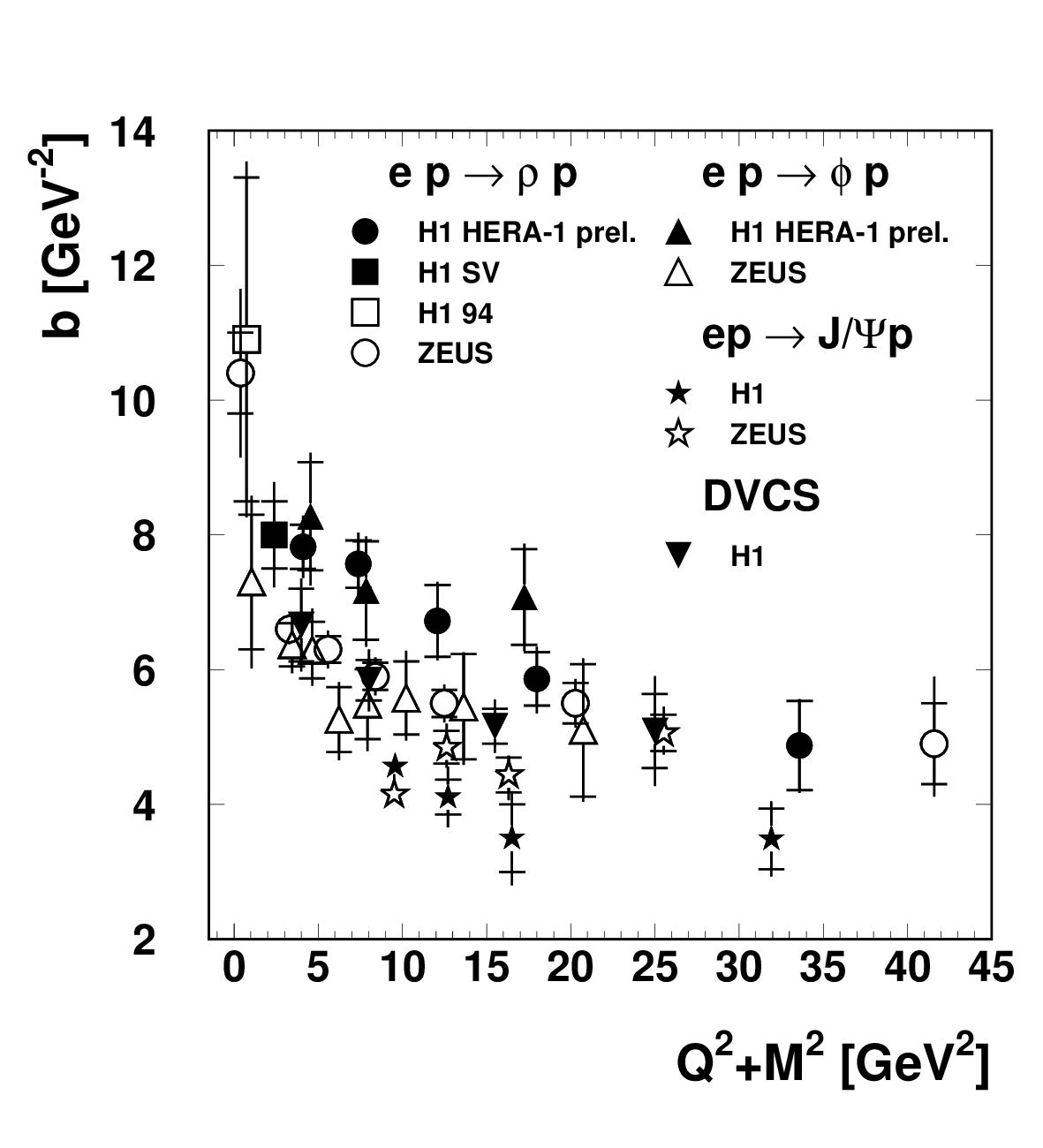}
\includegraphics[width=0.40\columnwidth]{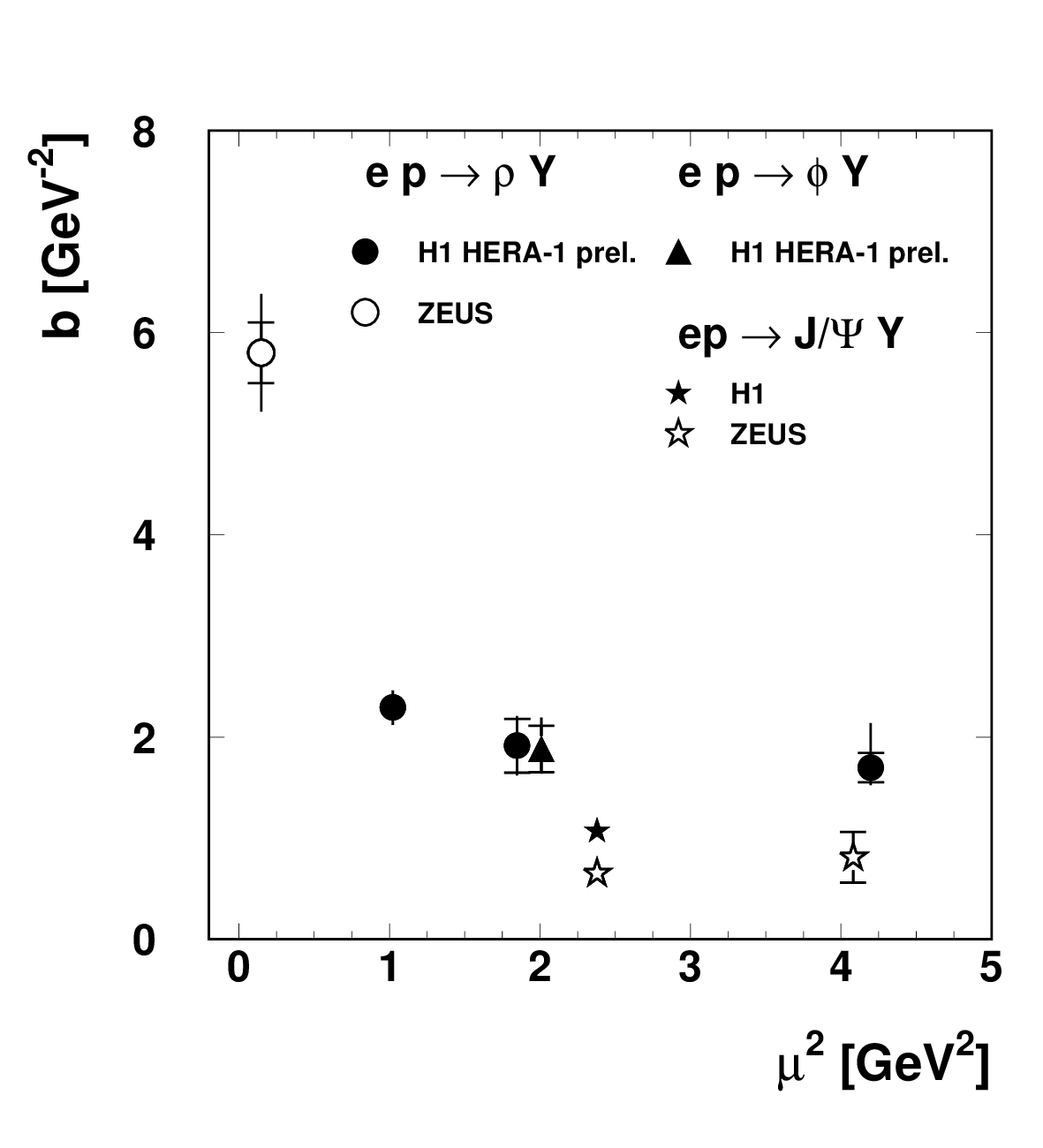}
\caption{Measurement of (left) the elastic and (right) the proton dissociative slopes $b$ of the 
exponential $t$ distributions, as a function of the scale 
$\mu^2 =  \qsqplmsq /4$ for VM production and $\mu^2 = \qsq$ for DVCS.}
\label{fig:bslope}
\end{center}
\end{figure}

For $\modt\ \lapprox\ 1-2~\gevsq$, the \modt\ distributions are exponentially falling 
with slopes $b$: $\rm {d} \sigma / \rm {d}t \propto e^{-b |t|}$.
In an optical model approach, the diffractive $b$ slope is given by the convolution 
of the transverse sizes of the interacting objects: $b = b_{q \bar q} + b_Y + b_{\pom}$,
with contributions of the $q \bar q$ dipole, of the diffractively scattered system
(the proton or the excited system $Y$) and of the exchange (``Pomeron") system.
Neglecting effects related to differences in the WF, 
universal $b$ slopes are thus expected for all VM with the same $q \bar q$ dipole sizes,
i.e. with the same values of the scale $\mu^2 = \qsqplmsq /4$.
Conversely, elastic and proton dissociative slopes are expected to differ for all VM production at 
the same scale by the same amount, $b_p - b_Y$.
Measurements of elastic and proton dissociative $b$ slopes for DVCS and VM production
are presented in Fig.~\ref{fig:bslope} as a function of the scale $\mu$~\footnote{
Differences between the H1 and ZEUS measurements for elastic scattering are due
to differences in background subtraction. The major effect is due to the subtraction of \rhop\ 
production by H1, a contribution evaluated to be negligible by ZEUS. 
Another difference concerns the values used for the $b$ slopes of the proton dissociative 
contamination.}.

For \jpsi\ elastic production, 
the $b$ slope is $\lapprox\ 4. 5~\gevsqm$, with no visible \qsq\ dependence.
This value may be related to the proton form factor~\cite{Kowalski:2006hc}.
For proton dissociation, the $b$ slope is below 1~\gevsqm, 
putting an upper limit to the transverse size of the exchange 
(with the assumption that $b_Y \simeq 0$ for proton dissociation).

At variance with \jpsi\ production, which is understood as a hard process already in 
photoproduction, a strong decrease of $b$ slopes for increasing values of 
$\mu^2 = \qsqplmsq /4$ is observed for light VM production, both in elastic and proton 
dissociative scattering. 
A similar scale dependence is observed for DVCS.
This is consistent with a shrinkage of the size of the initial state object with increasing \qsq,
i.e. in the VM case a shrinkage of the colour dipole.
It should however be noted that, both in elastic and proton dissociative scatterings,
$b$ slopes for light VM remain larger than for \jpsi\ when compared at the same 
values of the scale $\qsqplmsq /4 $ up to $\gapprox\ 5~\gevsq$.
The purely perturbative domain may thus require larger scale values.

\section{From soft to hard diffraction: {\boldmath $W$} dependences vs. mass and 
                  {\boldmath $Q^2$}}
                                                      \label{sec:W}

\begin{figure}[htbp]
\includegraphics[width=0.52\columnwidth]{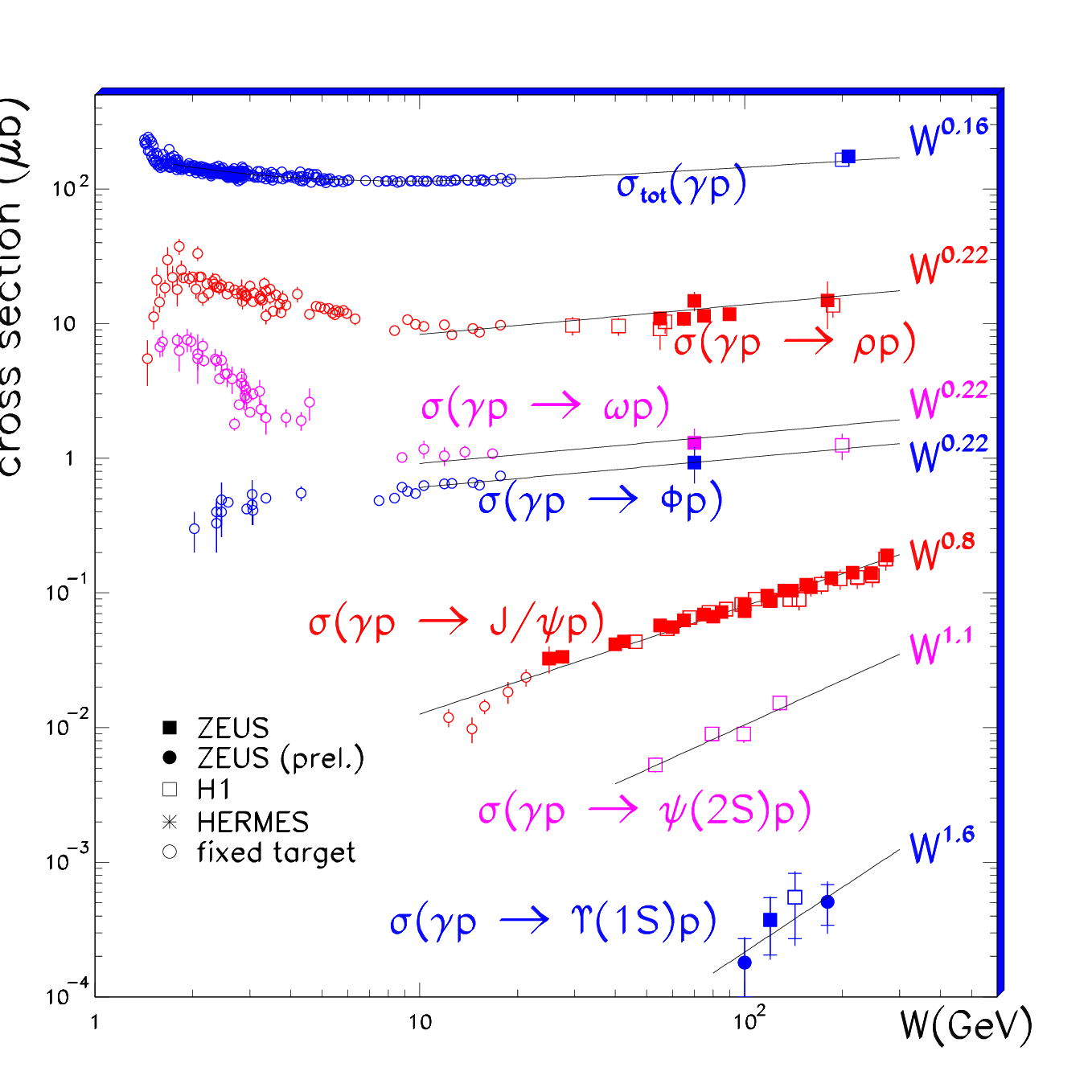}
\includegraphics[width=0.48\columnwidth]{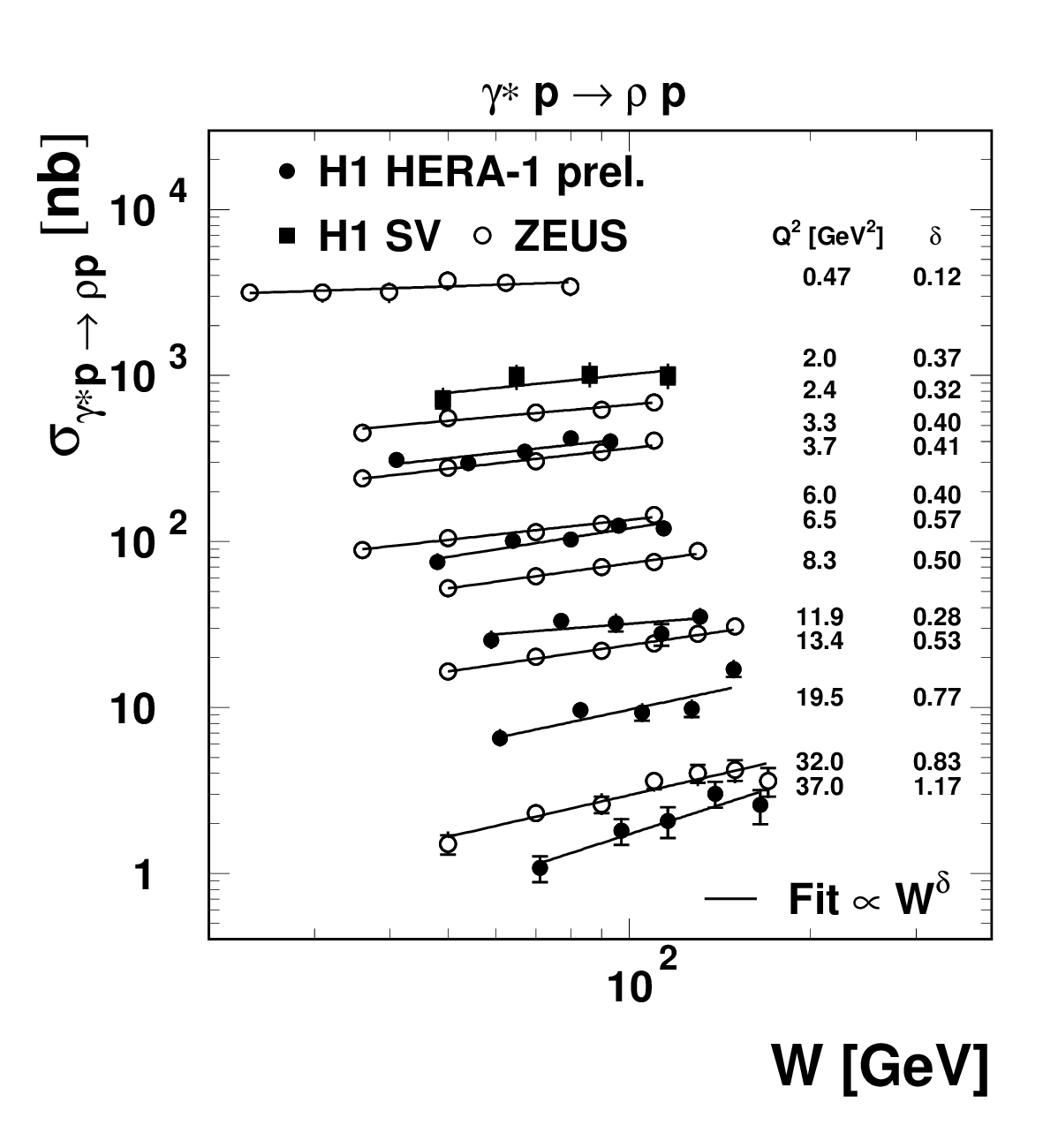}
\caption{ 
$W$ dependences of (left) total and VM photoproduction cross sections; 
(right) \rh\ electroproduction for several values of \qsq.
The lines show fits to the form $W^{\delta}$.}
\label{fig:sigmagp}
\end{figure}

Figure~\ref{fig:sigmagp}-left presents measurements as a function of $W$
of the total photoproduction cross section and of the exclusive 
photoproduction cross sections of several VM;
\rh\ electroproduction cross sections for several values of \qsq\ are shown in 
Fig.~\ref{fig:sigmagp}-right.
As expected for decreasing dipole sizes, the cross sections at fixed values of $W$ 
decrease significantly with increasing VM mass or \qsq.
In addition, different reactions exhibit strongly different $W$ dependences.
The total photoproduction cross section and the photoproduction of light VM 
show weak energy dependences, typical of soft, hadron--hadron processes.
In contrast, increasingly steep $W$ dependences are observed with increasing
mass or \qsq.
In detail, the $W$ dependences are investigated using a parameterisation inspired
by Regge theory, in the form of a power law with a linear parameterisation of the 
effective trajectory
\begin{equation}
\sigma \propto W^{\delta}, \ \ \ \ \
         \delta =  4 \ ( \alpom  - 1), \ \ \ \ \
                            \alpom(t) = \alpom(0) + \alp \cdot t.
                                                                       \label{eq:regge}
\end{equation}

\begin{figure}[htbp]
\begin{center}
\includegraphics[width=0.40\columnwidth]{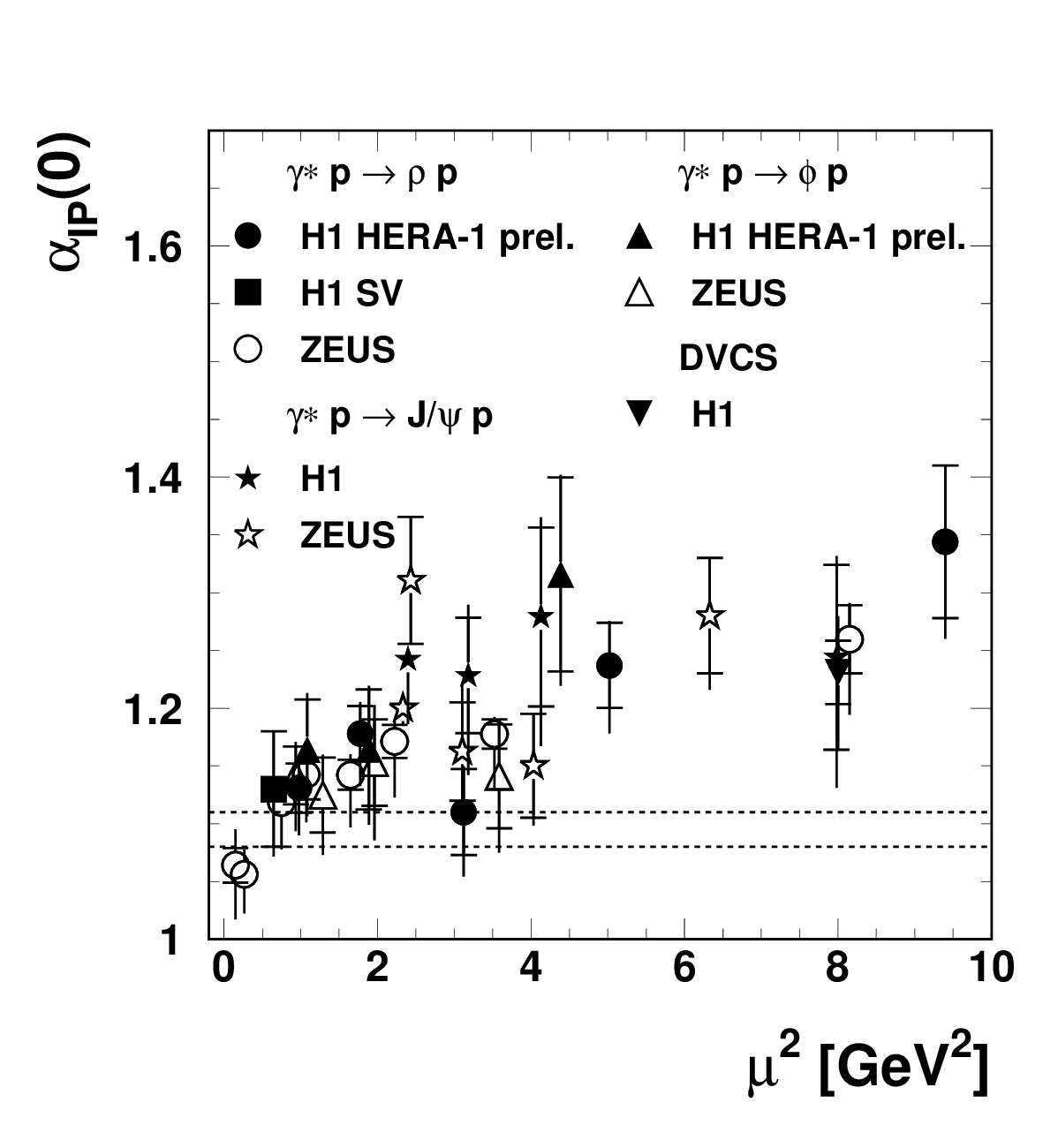}
\includegraphics[width=0.40\columnwidth]{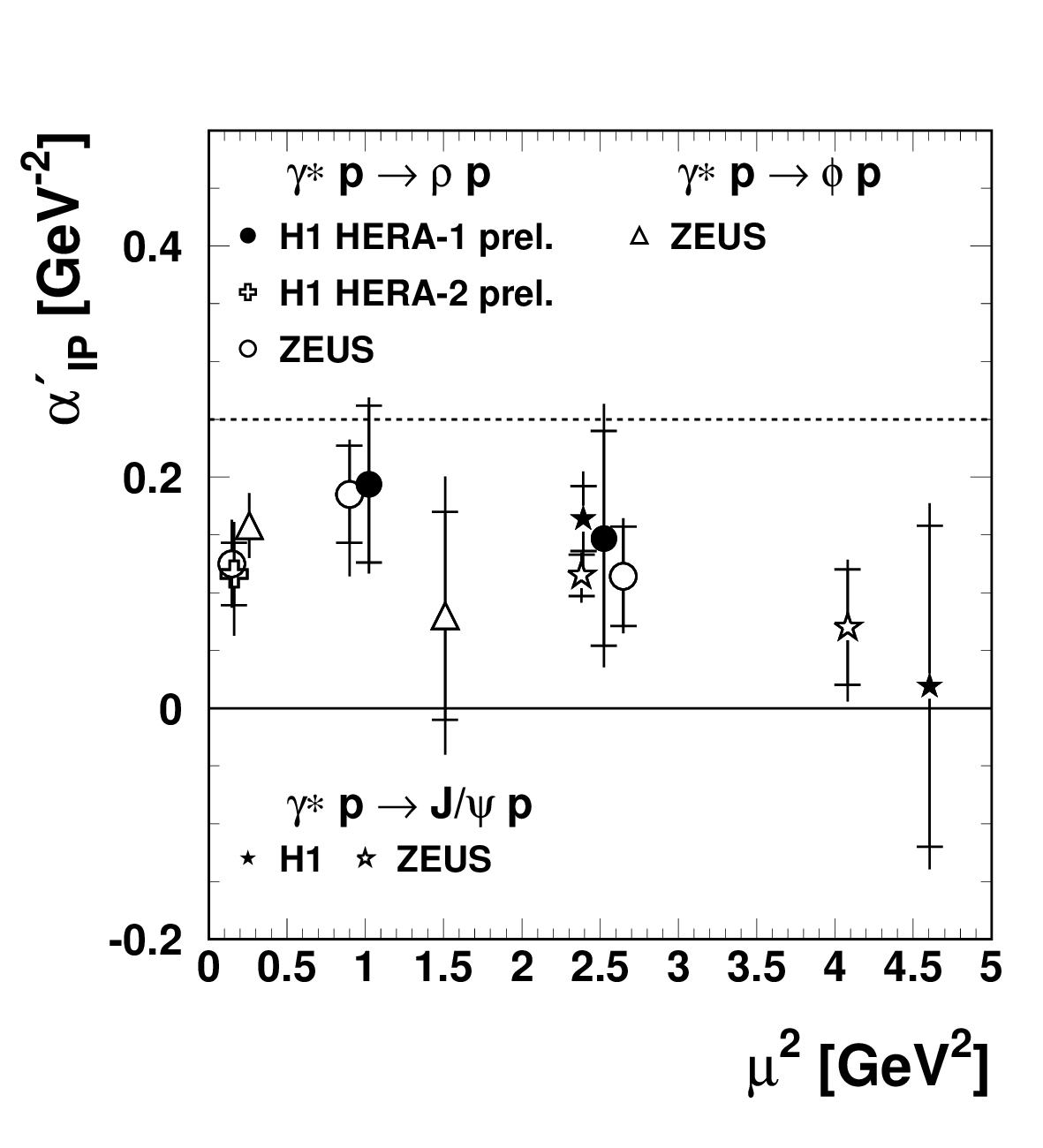}
\caption{
Values of (left) the intercept $\alpom(0)$ and (right) the slope \alp\ of the effective Pomeron 
trajectory, obtained 
from fits of the $W$ cross section dependences to the form 
$d\sigma/dt \propto W^{4(\alpha_{I\!\!P}(0) + \alpha^\prime \cdot t -1)}$.
The scales are $\mu^2 = \qsq$ for DVCS and $\mu^2 = \qsqplmsq /4 $ for VM production.
The dotted lines represent typical values for hadron--hadron scattering.}
\label{fig:trajectory}
\end{center}
\end{figure}

The intercept $\alpom(0)$ of the effective trajectory 
quantifies the energy dependence of the reaction for $t = 0$.
The evolution of $\alpom(0)$ with $\mu^2$ is shown in Fig.~\ref{fig:trajectory}-left.
Light VM production at small $\mu^2$ gives values of $\alpom(0)\ \lapprox\ 1.1$, similar 
to those measured for soft hadron--hadron interactions~\cite{dola,*cudell}.
In contrast larger values, $\alpom(0)\ \gapprox\ 1.2$, are observed for DVCS, for light VM 
at large \qsq\ and for heavy VM at all \qsq.
This increase is related to the large parton densities in the proton at small $x$, which are resolved 
in the presence of a hard scale:
the $W$ dependences of the cross section is governed by the hard $x^{-\lambda}$ 
evolution of the gluon distribution, with
$\lambda \simeq 0.2$ for $Q^2 \simeq M^2_{J/\psi}$.
The $W$ dependences of VM cross sections, measured for different $Q^2$ values, are 
reasonably well described by pQCD models (not shown).
In detail these are however sensitive to assumptions on the imput gluon densities in the 
domain $10^{-4}\ \lapprox\ x\ \lapprox\ 10^{-2}$ which is poorly constrained by inclusive 
data~\cite{Martin:2007sb,teubner}.

The slope \alp\ in eq.~(\ref{eq:regge}) describes the correlation between the $t$ and 
$W$ dependences of the cross section.
The measurement of the evolution with $t$ of the $\delta$ exponent 
can be parameterised as a $W$ dependence of the $b$ slopes,
with $b=b_0+4\alpha'\ln W/W_0$.
In hadron--hadron scattering, positive values of \alp\ are measured, with
$\alp \simeq 0.25~\gevsqm$~\cite{alphaprim1}.
This shrinkage of the diffractive peak indicates the expansion with energy of the size of
the interacting system, i.e. the expansion of the gluon cloud in the periphery of the 
interaction.
HERA measurements are presented in Fig.~\ref{fig:trajectory}-right.
The values of \alp\ are positive and appear smaller than in hadron--hadron interactions, 
also for \rh\ photoproduction.
This suggests a limited expansion of the systems considered here on the relevant 
interaction time scale.
In a BFKL approach, \alp\ is related to the average $k_t$ of gluons around the ladder in their 
random walk, and is expected to be small~\cite{Brodsky:1998kn}.


\section{ {\boldmath $Q^2$} dependences in DVCS and VM production }
                                                      \label{sec:qsq}

The description of the \qsq\ dependences of the cross sections is 
a challenge, in view of the presence of higher order corrections 
and of non-perturbative effects, especially for transverse VM production.

\subsection{DVCS}
                                                      \label{sec:qsq-DVCS}

The DVCS cross section depends on the proton GPD distributions.
To investigate the dynamical effects due to QCD evolution, the \qsq\ dependence has been
measured and studied~\cite{Aaron:2007cz} as a function of the dimensionless scaled variable 
$S$,
$$S = \sqrt { \sigma_{DVCS} \ Q^4 \ b(Q^2) \ / \ (1+\rho^2)},$$
which removes the effects of the photon propagator and of the \qsq\ dependence of the 
$b$ slope, 
and of the ratio $R$ of the imaginary parts of the DVCS and DIS amplitudes,
$$R =  \frac  { {\cal I}m {{ A}}(\gamma^* p \to \gamma  p)_{t=0}      }
                   { {\cal I}m {{ A}}(\gamma^* p \to \gamma^*  p)_{t=0}   }
    = 4  \frac { \sqrt{\pi \  \sigma_{DVCS} \  b(Q^2)}  }
                  { \sigma_T(\gamma^* p\rightarrow X)    }      \ \sqrt{1+\rho^2},$$
with $\sigma_T(\gamma^* p \rightarrow X) = 4\pi^2 \alpha_{EM} F_T(x,Q^2)/ Q^2$,
$F_T = F_2 - F_L$ and $\rho = {\cal R}e {{ A}} / {\cal I}m {{ A}}$ determined from 
dispersion relations~\cite{Favart:2003cu}.

\begin{figure}[htbp]
 \begin{center}
\includegraphics[width=0.50\columnwidth]{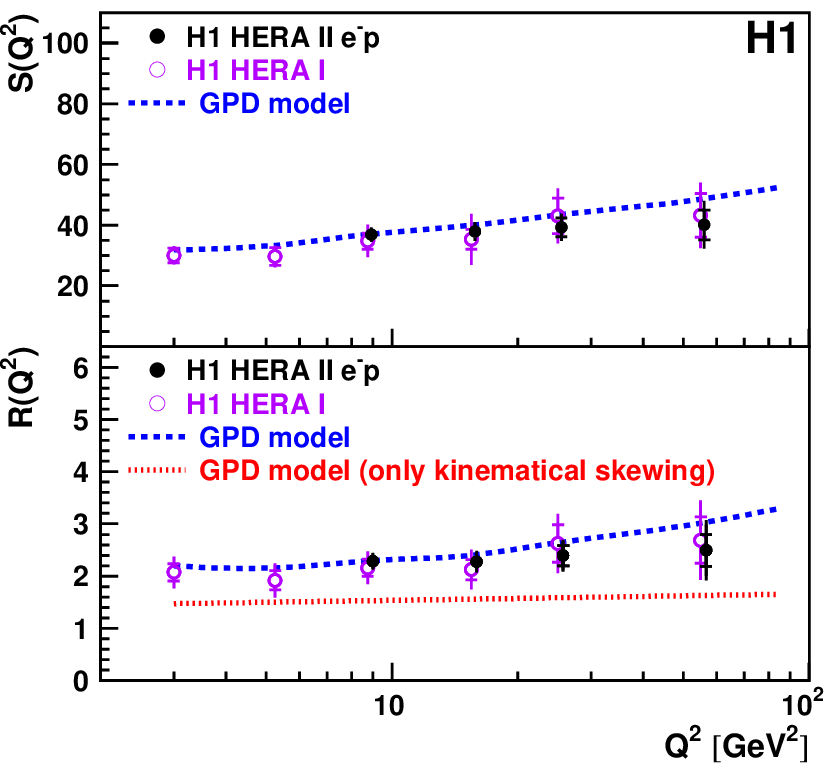}
\includegraphics[width=0.48\columnwidth]{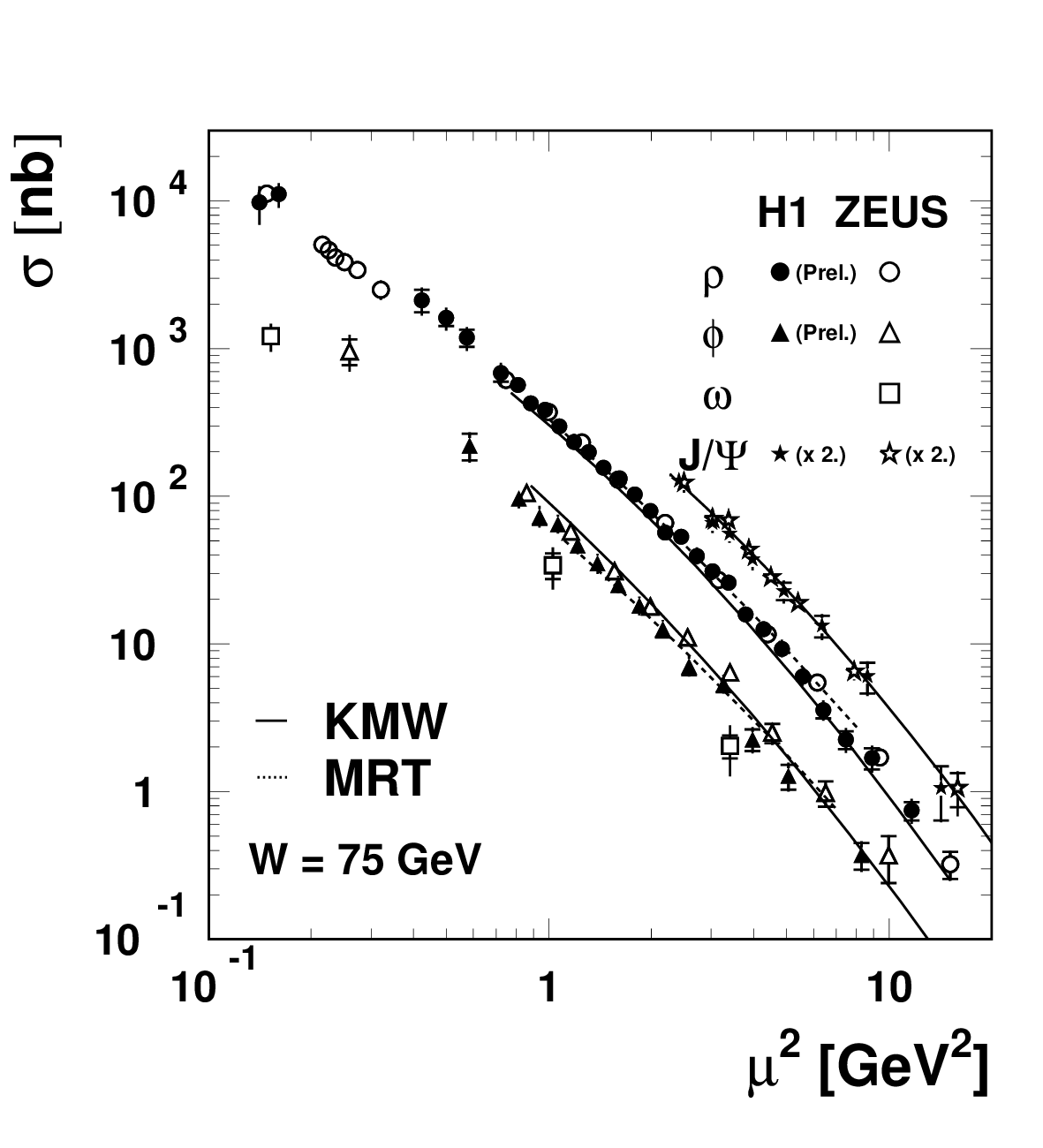}
 \end{center}
\caption{
(left) \qsq\ dependences of the observables $S$ and $R$ for DVCS (see text);
(right) \rh, \om, \ph\ and \jpsi\ elastic production cross sections, as a function
of the scale $\mu^2 = \qsqplmsq /4 $; for readability of the figure, the \jpsi\ cross sections
have been multiplied by a factor 2.
The curves are predictions of the KMW~\protect\cite{Kowalski:2006hc} and
MRT~\protect\cite{Martin:1996bp} models.}
\label{fig:qsq-dep}
\end{figure}

Figure~\ref{fig:qsq-dep}-upper-left shows a weak rise 
of $S$ with $Q^2$, which is reasonably well described by the GPD 
model~\cite{Freund:2003qs} using the CTEQ PDF parameterisation~\cite{Pumplin:2002vw}. 
The large effect of skewing is visible in Fig.~\ref{fig:qsq-dep}-lower-left, where 
the variable $R$ takes values around $2$, instead of $1$ in the absence of skewing.
GPD calculations~\cite{Freund:2003qs} compare well with measurements, whereas the same
figure shows that it is not sufficient to include only the kinematic contribution to skewing, 
and that the $Q^2$ evolution of the GPD must also be taken into account.

\subsection{Vector mesons}
                                                      \label{sec:qsq-VM}

The elastic production cross sections \rh, \om, \ph\ and \jpsi\  are shown in 
Fig.~\ref{fig:qsq-dep}-right, as a function of the scaling variable $\qsqplmsq /4$ (for readability,
the \jpsi\ cross sections have been multiplied by 2)~\footnote{Whereas the H1 and ZEUS 
measurements for \rh\ agree well, \ph\ measurements of ZEUS are a factor 1.20 above H1.
When an improved estimation of the proton-dissociation background, investigated for the latest 
ZEUS \rh\ production study~\cite{Chekanov:2007zr}, is used to subtract this background in 
their \ph\ analysis, 
the cross section ratio of the two experiments is reduced to 1.06, which is within experimental 
errors.}.
It is striking that, whereas light VM and \jpsi\ production cross sections for the same value of 
\qsq\ differ by orders of magnitude (see Fig.~\ref{fig:sigmagp}-left for $\qsq = 0$), they are close
when plotted as a function of the scaling variable $\qsqplmsq /4 $, up to the factors
accounting for the VM charge content ($\rho : \phi : J/\psi = 9:2:8$)~\footnote{For detailed 
comparisons, modifications due to WF effects, as observed in VM 
electronic decay widths, may need to be taken into account.}.
This supports the dipole approach of VM production at high energy.

The  cross sections are roughly described by power laws
$1 / (Q^2 \!+ \!M_V^2)^n$, with $n \simeq 2.2-2.5$.
The simple $n = 3$ dependence expected in a two-gluon approach for the dominant 
longitudinal cross sections is modified not only by an additional factor $1 / \qsq$ in the
transverse amplitudes, but also by the 
\qsq\ dependence of the gluon distribution at small $x$, described by the DGLAP evolution
equations.
Calculations using the $k_t$-unintegrated gluon distribution model of MRT~\cite{Martin:1996bp} 
or the GPD model~\cite{kroll} (not shown) give reasonable descriptions of the \qsqplmsq\ 
dependences.
However, in detail, a good description necessitates the precise modelisation of the \qsq\ 
dependence of the longitudinal to transverse cross section ratio $R$, with non-perturbative 
effects affecting $\sigma_T$.
Dipole models using different saturation and WF parameterisations, e.g. the 
FSS~\cite{Forshaw:2003ki}, KMW~\cite{Kowalski:2006hc} and DF~\cite{Dosch:2006kz} models, 
attempt at describing VM production over the full \qsq\ range, including photoproduction, 
with reasonable success.

\section{Matrix elements and \boldmath {$\sigma_L / \sigma_T$} }
                                                      \label{sec:matrix}

\begin{figure}[htbp]
\begin{center}
\hspace{-0.4cm}{\includegraphics[width=.25\columnwidth]{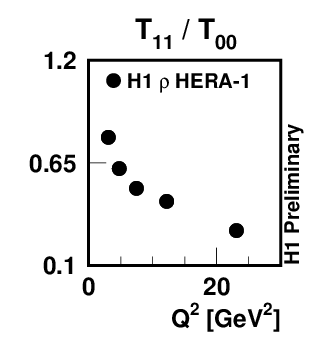}}
\hspace{-0.4cm}{\includegraphics[width=.25\columnwidth]{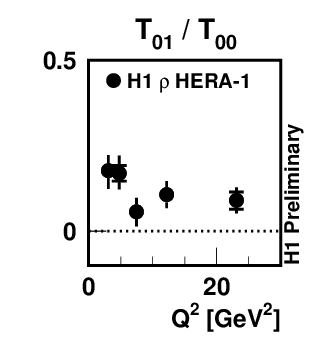}}
\hspace{-0.4cm}{\includegraphics[width=.25\columnwidth]{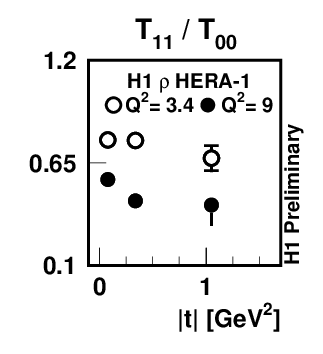}}
\hspace{-0.4cm}{\includegraphics[width=.25\columnwidth]{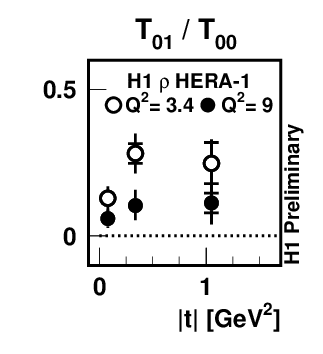}}
\caption{Amplitude ratios $T_{11} / T_{00}$ and $T_{01} / T_{00}$ as a function of \qsq\ 
and \modt\ (for two bins in \qsq ), for \rh\ electroproduction.
The dotted lines represent the SCHC approximation.}
\label{fig:amplitudes}
\end{center}
\end{figure}

\begin{figure}[htbp]
\begin{center}
\includegraphics[width=0.46\columnwidth]{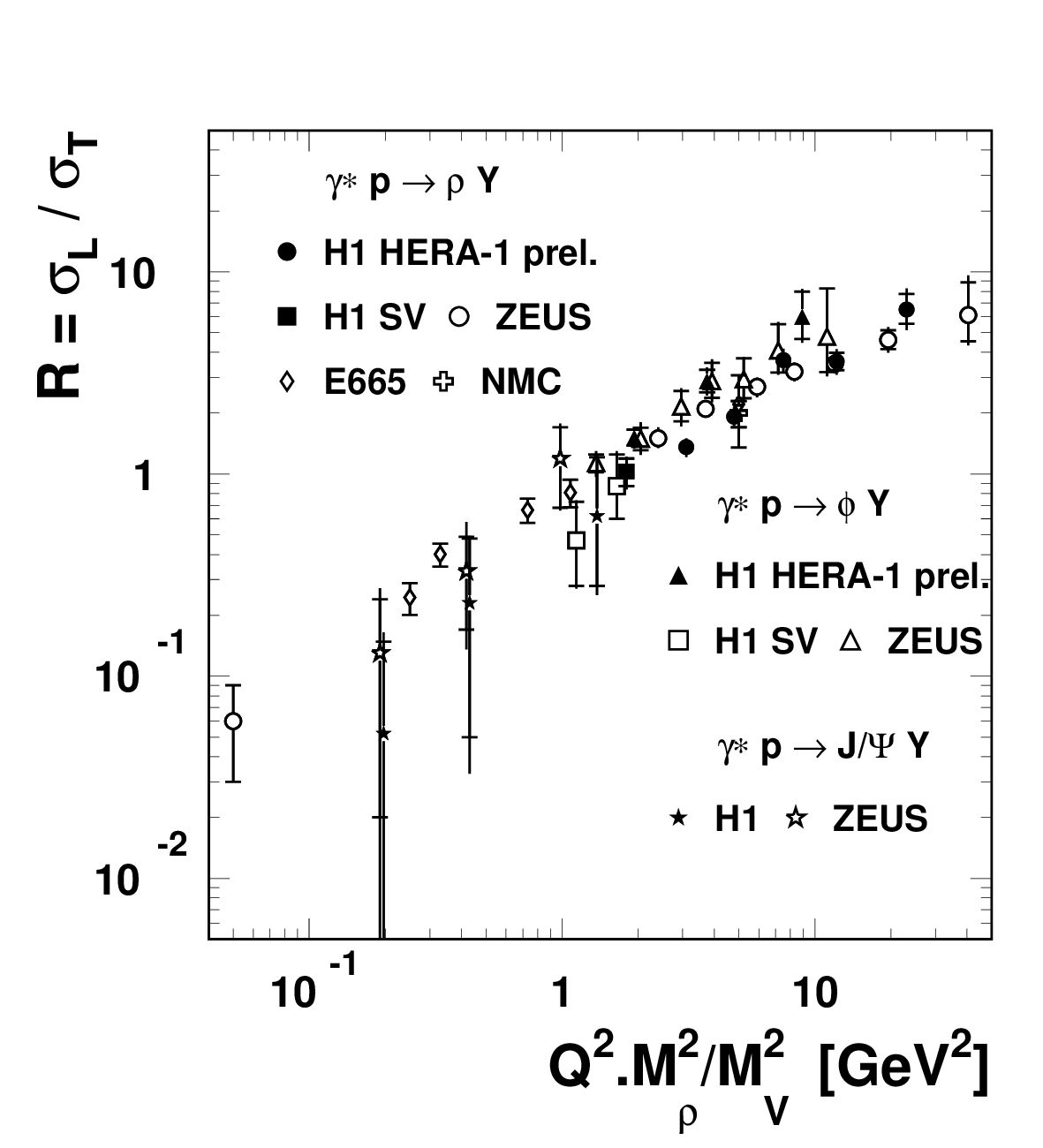}
\includegraphics[width=0.50\columnwidth]{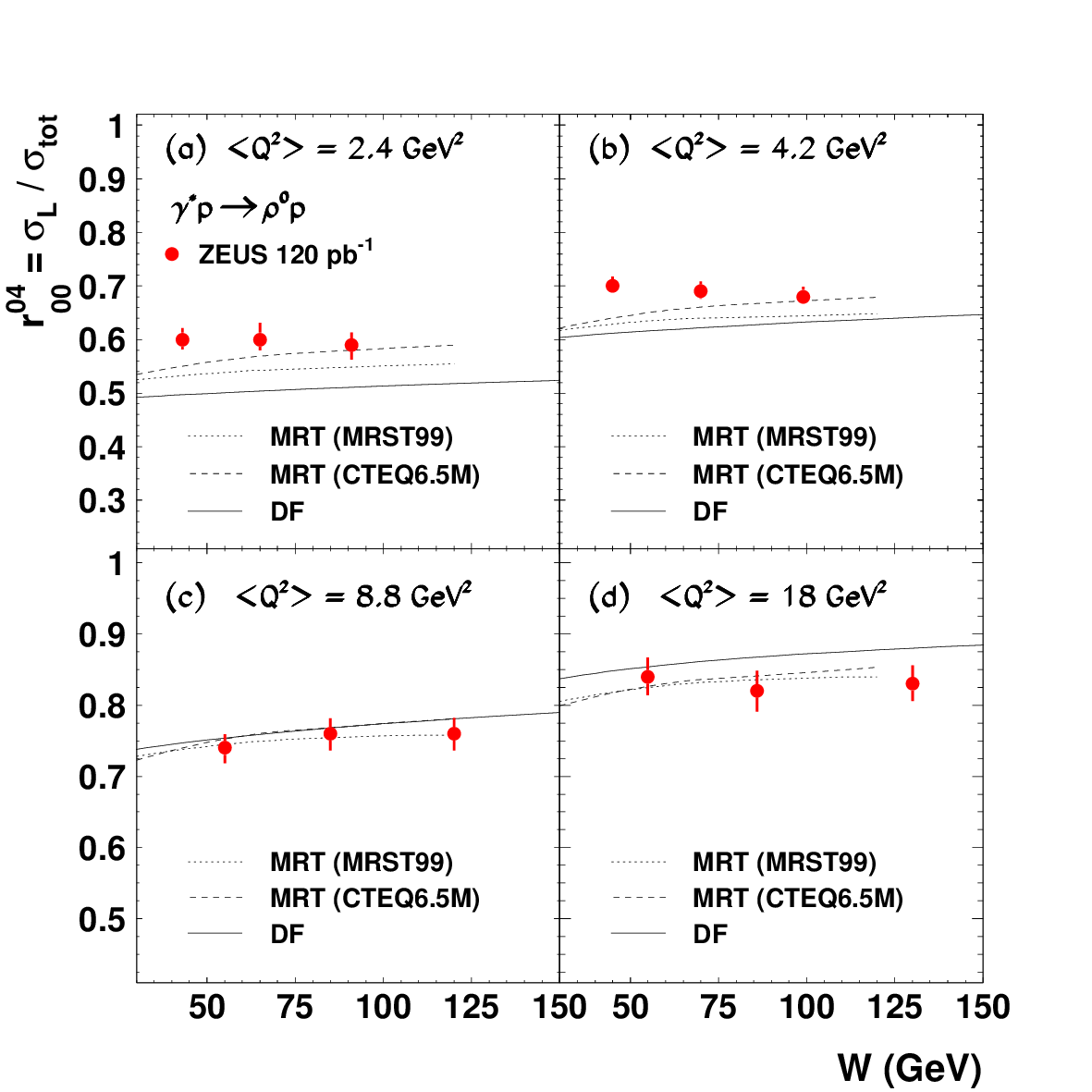}
\caption{
Cross section ratio $R=\sigma_L/\sigma_T$ as a function of 
(left) the scaling variable $Q^2 \cdot M^2_{\rho} / M^2_{V}$ for different VM; 
(right) the centre of mass energy $W$ in several \qsq\ bins for \rh\ electroproduction,
compared to model predictions.}
\label{fig:R}
\end{center}
\end{figure}

Measurements of  the VM production and decay angular distributions give access to spin 
density matrix elements, which are related to the helicity amplitudes 
$T_{\lambda_{V} \lambda_{\gamma}}$~\cite{sch-w}.
Analyses of \rh, \ph\ and \jpsi\ photo- and electroproduction indicate the dominance of
the two $s$-channel helicity conserving (SCHC) amplitudes, the transverse $T_{11}$ 
and the longitudinal $T_{00}$ amplitudes, 
In the accessible \qsq\ ranges, \jpsi\ production is mostly transverse, whereas for light VM 
electroproduction the longitudinal amplitude $T_{00}$ dominates (see Fig.~\ref{fig:amplitudes}a 
and Fig.~\ref{fig:R}a).
In \rh\ and \ph\ electroproduction, a significant contribution of the transverse to longitudinal 
helicity flip amplitude \tzu\ is observed.
The amplitude ratio $T_{01} / T_{00}$ decreases with \qsq\ (Fig.~\ref{fig:amplitudes}b) and increases 
with \modt\ (Fig.~\ref{fig:amplitudes}d), as expected (see e.g.~\cite{Ivanov:1998gk});
the SCHC amplitude ratio $T_{11} / T_{00}$ decreases with \modt\ (Fig.~\ref{fig:amplitudes}c) .

Figure~\ref{fig:R} presents measurements of the longitudinal to transverse cross ratio 
$R = \sigma_L / \sigma_T \simeq |T_{00}|^2 / |T_{11}|^2$ (in the SCHC approximation).
The behaviour $R \propto \qsq / \msq$ predicted for two-gluon exchange is qualitatively 
verified for all VM production, in fixed target and HERA experiments. 
This is shown in Fig.~\ref{fig:R}-left, where $R$ is plotted as a function of the scaled variable  
$Q^2\cdot M^2_{\rho}/M^2_{V}$.
However, the \qsq\ dependence is tamed at large values of \qsq, a feature which 
is expected and relatively well described by pQCD based calculations, e.g. the GPD 
model~\cite{kroll}, the $k_t$-unintegrated models~\cite{Martin:1996bp,Ivanov:1998gk}
or the dipole model~\cite{Kowalski:2006hc}.

The cross section ratio $R$ for \rh\ electroproduction is also found to depend very 
significantly on the dipion mass $M_{\pi\pi}$ (not shown),
in line with the $\qsq / \msq$ dependence if the relevant mass is the dipion mass 
rather than the nominal \rh\ resonance mass.
Following the  MRT model approach~\cite{Martin:1996bp}, this suggests a limited influence
of the WF on VM production.

Figure~\ref{fig:R}-right shows that no strong dependence of $R$ with $W$ is observed.
Since transverse amplitudes are expected to include significant contributions of 
large dipoles, with a soft energy dependence, this suggests that large dipoles are also 
present in longitudinal amplitudes, due to finite size effects, i.e. a smearing of $z$ away from 
$z = 1/2$.
On the other hand, in the domain $\qsq\ \gapprox\ 10-20~\gevsq$, no strong dependence 
of $R$ with $W$ is expected from models.
It should also be noted that a significant phase difference is observed between the two 
dominant amplitudes, \tzz\ and \tuu~\cite{janssen2008}.
This indicates a difference between the ratios of the real to imaginary parts of the forward 
amplitudes.
Since these ratios are given by $\log 1/x$ derivatives of the amplitudes, the phase difference 
is an indication of different $W$ dependences.

\section{Large {\boldmath \modt}; BFKL evolution}
                                                       \label{sec:large t}

Large values of the momentum transfer \modt\ provide a hard scale for diffractive processes 
in QCD, with the dominance of the proton dissociative channel for $|t|\ \gapprox\ 1~\gevsq$.
It should be noted that for large \modt\ production, a hard scale is present at both ends of the 
exchanged gluon ladder.
No strong $k_t$ ordering is thus expected, which is typical for BFKL evolutions for sufficiently
high \modt\ values.
This is at variance with large \qsq\ VM production at low \modt, where a large scale is present 
at the upper
(photon) end of the ladder and a small scale at the proton end, implying that these processes 
are expected to be described by DGLAP evolutions, with strong $k_t$ ordering along the 
ladder.

\begin{figure}[htbp]
\begin{center}
\includegraphics[width=0.37\textwidth]{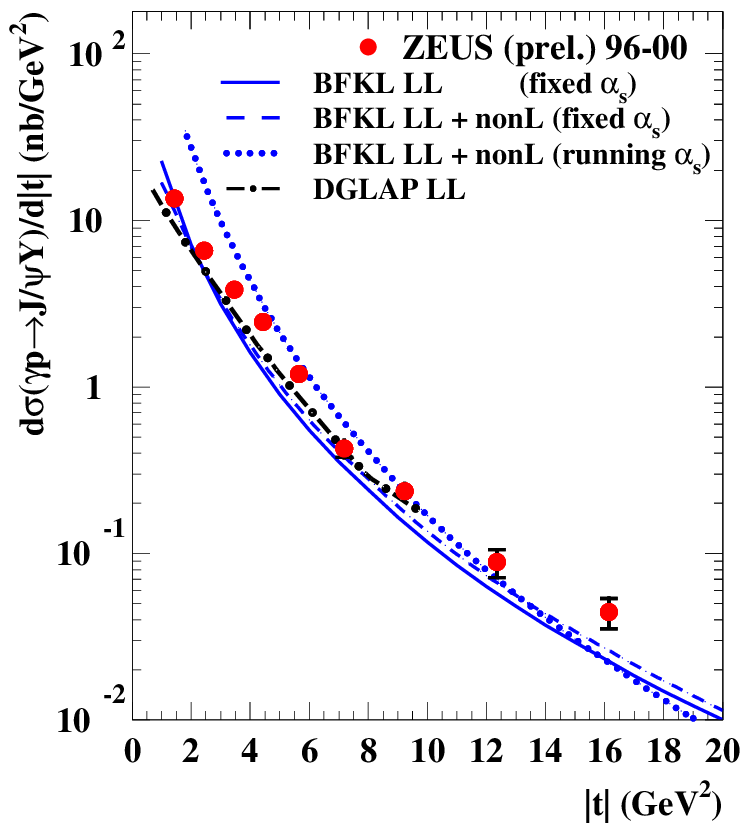}
\includegraphics[width=0.40\textwidth]{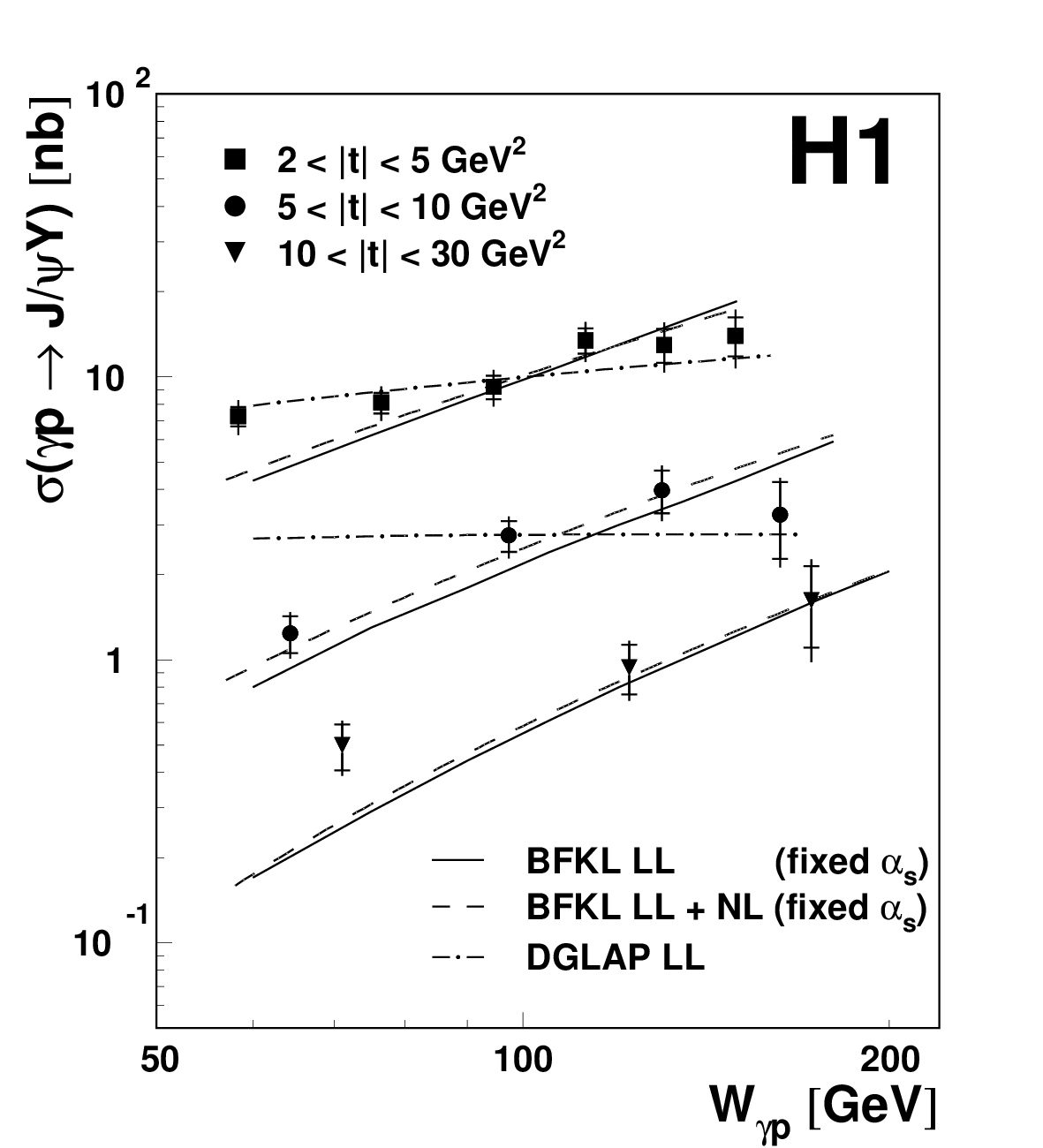}
\caption{
$t$ (left) and $W$ (right) dependences of \jpsi\ production with $\modt > 2~\gevsq$,
with comparisons to pQCD model predictions.}
\label{fig:bfkl}
\end{center}
\end{figure}

For \modt\ larger than a few~\gevsq, the $t$ dependences of the cross sections follow power laws, 
both for $\rho$~\cite{hight_zeus,*Aktas:2006qs} and $J/\psi$~\cite{Aktas:2003zi,*hight_jpsi_zeus} 
photoproduction. 
As shown by Fig.~\ref{fig:bfkl}-left, they are well described by pQCD calculations based on the 
BFKL equations with fixed $\alpha_s$~\cite{Enberg:2002zy}; predictions using the DGLAP 
evolution~\cite{Gotsman:2001ne} also describe the \jpsi\ data for $|t|\ \lapprox\ m_\psi^2$.
BFKL calculations describe the $W$ evolution (Fig.~\ref{fig:bfkl}-right), at variance with DGLAP, but 
do not describe well the spin density matrix elements.
For \rh, \ph\ and \jpsi\ photoproduction with $\modt\ \gapprox\ 2~\gevsq$, the slope \alp\ of the 
effective Regge trajectory tends to be slightly negative, but are compatible with 0.

\section{Conclusions}
                                                      \label{sec:concl}

In conclusion, studies of VM production and DVCS at HERA provide a rich and varied
field for the understanding of QCD and the testing of perturbative calculations over a 
large kinematical domain, covering the transition from the non-perturbative to the perturbative 
domain.
Whereas soft diffraction, similar to hadronic interactions, dominates light VM photoproduction,
typical features of hard diffraction, in particular hard $W$ dependences, show up
with the developments of hard scales provided by \qsq, the quark mass or \modt. 
The size of the interaction is accessed through the $t$ dependences.
Calculations based on pQCD, notably using $k_t$-unintegrated gluon distributions and GPD 
approaches, and predictions based on models invoking universal dipole--proton cross sections 
describe the data relatively well.
The measurement of spin density matrix elements gives a detailed
access to the polarisation amplitudes, which is also understood in QCD.
Large \modt\ VM production supports BFKL calculations.


\section*{Acknowledgements}

It is a pleasure to thank the numerous colleagues from H1 and ZEUS who contributed to the 
extraction of the beautiful data presented here, and to thank the theorist teams whose efforts 
led to a continuous increase in the understanding of diffraction in terms of QCD, the theory of 
strong interactions.
Special thanks are due to  L.~Motyka, T.~Teubner and G.~Watt for providing calculations 
used for the present review and to J.-R.~Cudell and L.~Favart for useful discussions and 
comments.
It is also a pleasure to thank the organisors of the HERA-LHC Workshop for the lively discussions 
and the pleasant atmosphere of the workshop.

\bibliographystyle{heralhc-0-1} 
{\raggedright
\bibliography{heralhc-0-1}
}

\end{document}